\documentclass[namedreferences]{SolarPhysics}
\usepackage[optionalrh]{spr-sola-addons} 
\usepackage{graphicx}        
\usepackage{color}           
\usepackage{url}             

\begin{document}

\begin{article}

\begin{opening}

\title{Can We Constrain the Solar Interior Physics Studying the Gravity-Mode Asymptotic Signature?}

\author{R.A.~\surname{Garc\'\i a}$^{1}$\sep
		S.~\surname{Mathur}$^{1}$\sep
             J.~\surname{Ballot}$^{2}$	
          }
\runningauthor{R.A. Garc\'\i a {\it et al.}}
\runningtitle{Can We Constrain the Solar Interior Physics Studying the  {\bf \emph{g}}-Mode Asymptotic Signature?}

   \institute{$^{1}$ 
Laboratoire AIM, CEA/DSM-CNRS - U. Paris Diderot - IRFU/SAp, 91191 Gif-sur-Yvette Cedex, France \url{rgarcia@cea.fr, smathur@cea.fr}\\ 
              $^{2}$Max-Planck-Institut f\"ur Astrophysik, Karl-Schwarzschild-Strasse 1, 85748 Garching, Germany \url{jballot@mpa-Garching.mpg.de}\\
                      }
\date{Received 18 October 2007; Accepted 28 February 2008}

\begin{abstract}

Gravity modes are the best probes to infer the properties of the solar radiative zone that represents 98$\%$ of the Sun's total mass. It  is usually assumed that high-frequency $g$ modes give information about the structure of the solar interior whereas low-frequency $g$ modes are more sensitive to the solar dynamics (the internal rotation).  In this work, we develop a new methodology, based on the analysis of the almost constant separation of the dipole gravity modes, to introduce new constraints on the solar models. To validate this analysis procedure, several solar models -- including different physical processes and either old or new chemical abundances (from, respectively, Grevesse and Noels (\emph{Origin and evolution of the elements}. Cambridge, England, 199, 15, 1993) and  Asplund, Grevesse, and Sauval (\emph{Astron. Soc. Pac. CS, San Francisco}, {\bf 336}, 25, 2005)) -- have been compared to another model used as a reference. The analysis clearly shows that this methodology has enough sensitivity to distinguish between some of the models, in particular, between those with  different compositions. The comparison of the models with the $g$-mode asymptotic signature detected in GOLF data favors the ones with old abundances. Therefore, the physics of the core -- through the analysis of the $g$-mode properties -- is in agreement with the results obtained in the previous studies based on the acoustic modes, which are mostly sensitive to more external layers of the Sun.

\end{abstract}
\keywords{Helioseismology, Observations; Interior, Radiative zone, Core}
\end{opening}

\section{Introduction}
The combination of detailed modeling and helioseismic observations has enabled us to accurately describe the solar interior from the surface to its center where the nuclear reactions take place. The precise characterization of the resonant acoustic (pressure or $p$) modes allowed constraining the theories describing the solar interior. The profiles of sound speed, density (\emph{e.g.} \opencite{BasJCD1997}; \opencite{CouSTC2003}), and differential rotation  \cite{ThoToo1996,HowJCD2000,ChaEls2001,CouGar2003}, as well as the position of the base of the convective zone (\emph{e.g.} \opencite{JCDGou1991}), and the prediction of the neutrino fluxes (\emph {e.g.} \opencite{STCCou2001}) are some examples of the constraints provided by the $p$-mode analyses.

Unfortunately, only a few $p$ modes propagate inside the inner core -- below 0.25 solar radius ($R_\odot$) -- containing about 50$\%$ of the solar mass. To progress in the knowledge of these layers, another type of mode is needed: the gravity ($g$) modes. These modes have not yet been unambiguously detected individually in the Sun (for example see \opencite{AppFro2000}; \opencite{GabBau2002}; \opencite{2006soho...18E..22E}) because they become evanescent in the convection zone reaching the surface of the Sun with very small amplitudes \cite{And1996,KumQua1996}. Some candidates have been detected \cite{STCGar2004,2007ApJ...668..594M} analyzing helioseismic data from the Global Oscillations at Low Frequencies (GOLF \cite{GabGre1995}) instrument on board the ESA/NASA \emph{Solar and Heliospheric Observatory} (SOHO \cite{DomFle1995}) spacecraft. These candidates have provided some scenarios about the possible dynamics of the solar core but it has been impossible to correctly label these peaks in terms of their orders and degrees ($n$, $\ell$, $m$) and to give the central frequencies of the candidates. Recently \inlinecite{2007Sci...316.1591G} were able to uncover the signature of the asymptotic properties of the dipole ($\ell=1$) $g$ modes measuring their constant spacing $\Delta P_1$. Indeed, for a given degree $\ell$, the difference $\Delta P_\ell$ of the periods of $g$ modes with consecutive radial order $n$ is almost constant when $n\gg\ell$ \cite{1980ApJS...43..469T,1986A&A...165..218P}. The comparison of the observations with some solar models -- including different rotation rates of the core -- showed the great sensitivity of this analysis to the dynamics of the solar core and it implied a rotation rate in the core faster than in the rest of the radiative region.  

During the last few decades, many new physical processes have been studied and included in the solar models while others have been refined in order to better match the helioseismic observations. Today, these models include better descriptions of the microscopic physics such as opacities, equation of state, nuclear reaction rates (\emph{e.g.} \opencite{1997A&A...327..349M} and references therein), microscopic diffusion \cite{1993ASPC...40..246M} as well as other processes, for example turbulence mixing in the tachocline \cite{1999ApJ...525.1032B}.

New studies of the solar spectrum using the most recent advances in 3D modeling of the solar atmosphere and taking into account Non Local Thermodynamic Equilibrium (NLTE) effects have provided new estimations of the solar photospheric composition \cite{2005ASPC..336...25A}. In comparison with the older estimations (\emph{e.g.} \opencite{1993oee..conf...14G}) based on 1D models and the LTE approximation, this new measurement leads to a smaller metallicity and to a substantially modification of the abundances of C, N, and O, which are crucial in solar modeling. The standard models computed with these new abundances yield a larger discrepancy with the helioseismic observations.
They disagree with sound-speed and density profiles inferred from acoustic modes (for example see \opencite{2004PhRvL..93u1102T}; \opencite{2005ApJ...621L..85B}; \opencite{2005ApJ...627.1049G}) and we need, for example, to modify some opacities to reduce the discrepancy. It has also been shown that the models computed with these new abundances modify the frequencies of the resonant low-degree $p$ and $g$ modes \cite{2007ApJ...655..660B,2007ApJ...670..872C,2007A&A...469.1145Z,2007ApJ...668..594M}.

In this paper we use a set of solar models (briefly described in Section 2) -- including different physical processes combined with the old and new abundances of the Sun and with different core rotation rates --  that we utilize as a tool to study whether the methodology used to detect the dipole gravity modes in the Sun by \inlinecite{2007Sci...316.1591G} (and summarized in Section 3) is able to distinguish between these models. To do so, we  compare in Section 4  all of them to our reference model: the Saclay seismic model \cite{STCCou2001}. Finally, once the method has been demonstrated, we compare, in Section 5, the solar models with the GOLF data.

\section{Description of the Solar Models}

Several solar models have been computed with the Code d'\'Evolution Stellaire Adaptatif et Modulaire (CESAM) \cite{1997A&AS..124..597M}, which differ from each other by the physical processes that they include. They have been developed to study the influence of different physical processes and the impact of the new abundances on the $g$-mode frequencies. These models are fully explained in \inlinecite{2007ApJ...668..594M}. Here we recall briefly their main characteristics. 

First let us describe the solar model that has been taken as a reference: the so-called seismic model (corresponding to the model $seismic_2$ of \opencite{CouSTC2003}). It is important to note that this model is produced by an evolutionnary code and is not directly deduced from a seismic inversion. It is ``seismic'' because some physical quantities, such as opacities or nuclear reaction rates, have been tuned to better reproduce the sound-speed profile in the radiative zone deduced from the analysis of  $p$-mode observations. One of its aims was to improve the prediction of the neutrino flux. This model includes turbulent diffusion in the tachocline and the chemical composition of \inlinecite{1993oee..conf...14G}.

As we want to compare the impact of different physical processes in our work, five other solar models have been used. They mainly differ in three physical inputs: the microscopic diffusion (including the gravitational settling), the horizontal turbulent diffusion in the tachocline, and the abundances of chemical elements. The model called \emph{no diffusion} does not treat any diffusion processes, whereas all of the following models include it through the prescription of \inlinecite{1993ASPC...40..246M}. Although today this model is obsolete, it is a good test for the detection methodology, since the $g$-mode frequencies are very different as compared to the seismic ones and therefore, the analysis should be able to clearly separate them.

The second process, the tachocline mixing, is introduced in the code following the prescription of \inlinecite{1992A&A...265..106S}. The models including this process are called \emph{tacho} while the ones without this turbulent effect are named \emph{std}. This turbulent horizontal motion contributes to the thinness of the tachocline responsible for the mixing of chemical elements. Finally, the chemical composition is either taken from \inlinecite{1993oee..conf...14G} (\emph{GN93} hereafter) or from the recent release of \opencite{2005ASPC..336...25A} (\emph{As05} hereafter). The difference between these two chemical abundances is a decrease of the heavy element by $\approx$ 30\%. Such differences substantially affect, for instance, the Rossland opacities and lead to noticeably different thermal stratifications. The models using the old abundances have the suffix \emph{93} whereas the models including the new abundances have the suffix \emph{05}. Table~\ref{tab1} summarizes the main characteristics of all of the models used. Finally, to use a model calculated with another stellar evolution code, the following study is also done with model S from \inlinecite{JCDDap1996}, which is a standard model including the microscopic diffusion and using \emph{GN93}. 

\begin{table}[htb*]
\caption{The main characteristics of the solar evolution models used to compute the $g$-mode frequencies.}
\label{tab1}
\begin{tabular}{lccccc}
\hline
 & Microscopic & Diffusion in &  &  \\ 
Solar model\hspace*{3em} &diffusion & tachocline&\emph{GN93} & \emph{As05}& \\
 \hline
 no diffusion\dotfill & & & x & \\ 
 std93\dotfill & x & & x & \\ 
 std05\dotfill & x & & & x \\ 
 tacho93\dotfill & x & x & x &  \\ 
 tacho05\dotfill & x & x & & x   \\ 
 \hline
\end{tabular}
\end{table}%

From each stellar structure, eigenfrequencies have been computed with the adiabatic pulsation package (\url{adipack} provided by J. Christensen-Dalsgaard, available on \url{http://www.phys.au.dk/~jcd/adipack.n/}). The computation of $g$-mode frequencies can become very sensitive to the numerical approach, especially to the treatment of the boundary condition at the center. Thereby, the numerical uncertainties on the computed $g$-mode periods become large at very low frequency and we should be extremely cautious. However, in the range used for this study (between 2 and 8.5 hours, see Section~3), it has been shown that, for a given model, frequencies are modified by less than 50~nHz by changing the numerics (see \opencite{2007MathurPhD}), leading to a maximum uncertainty around 50~seconds on the mode periods. Such errors are smaller than the observed deviations between the different models used here. In fact, they are  too small (by more than an order of magnitude) to be correctly distinguished by the methodology developed in this paper.

\begin{figure}[htb*]  
 \centerline{\includegraphics[width=0.8\textwidth,clip=]{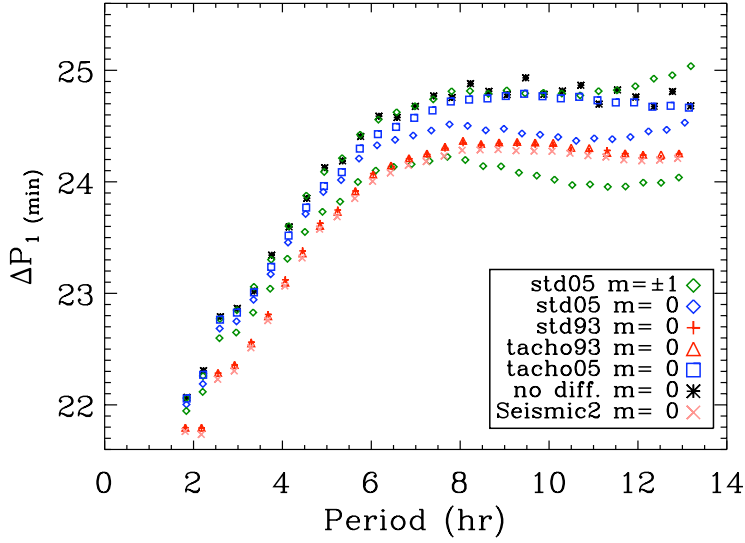} }
              \caption{Separations in period ($\Delta P_1$) of the $m$=0 components of the $g$ modes computed using the models described in Table~\ref{tab1}. For clarity, we have plotted the $m \pm$1 components only for one model: the {\it std05} (with a solid rotation of $\Omega_{rz}= 433$ nHz in the radiative zone 
 and a rotation in the convection zone that reproduces the inferred profile deduced from helioseismology data).
 }
   \label{DP1}
   \end{figure}

It is usually considered that the high-frequency $g$ modes (above 150 $\mu$Hz) are very sensitive to the physical processes included in the models (the structure) giving up to 5 $\mu$Hz of difference between the models, while at lower frequencies -- inside the asymptotic region --  the frequency differences are smaller than a fraction of a $\mu$Hz and they could be more useful to study the dynamics of the solar interior (for example see \opencite{2007ApJ...668..594M}). Indeed, we have already seen that the difference of the periods of two consecutive $g$ modes with the same degree ($\ell$) and successive radial order ($n$) is almost constant. In Figure~\ref{DP1} the difference ($\Delta P_1$) of the central frequencies ($m$=0 components) of the $g$ modes have been plotted for all the models described in Table~\ref{tab1}. The asymptotic regime is reached for periods above $\approx $ five hours and the differences between all of the models considered here are within one minute (from 24 to 25 minutes). Thus, it is going to be very difficult to distinguish between all these models.

Moreover, until now we have only considered the effect of the structure on the $g$-mode frequencies but we cannot neglect the effect of the rotation of the solar interior on the other $m$-components of the modes. Indeed, in Figure~\ref{DP1} we have also drawn the $m\pm$1 components for one of the solar models, the {\it std05}, considering a core rotating as a rigid body, like the rest of the radiative region. The dispersion in $\Delta P_1$ introduced by the dynamics (without even considering a rapid rotation in the core, just a rigid rotation profile) is already of the same order of the structural effects due to the different physical processes and abundances considered. That means that both effects cannot be easily disentangled and we have to study both at the same time.
 
At low frequency, it seems more appropriate to study the $g$ modes in period, and not in frequency. 
Depending on the physical processes taken into account in the solar models, there can be important differences between the periods of the $g$ modes. Thus, we could expect to use these low-frequency $g$ modes to diagnose some physical processes or abundances taken into account in the models.


To complete this study we need to model the dynamics --the rotation rate-- of the solar interior, in particular, in the regions where the $g$ modes are trapped, \emph{i.e.} the radiative region. Today, the solar rotation rate is known well down to the solar core \cite{ThoJCD2003} but it is still uncertain in the deepest layers. We have thus computed the $g$-mode splittings corresponding to ten artificial rotation profiles $\Omega(r,\theta)$. These artificial profiles have a differential rotation in the convection zone (that mimics the real behavior), and a rigid rotation from 0.7 down to a given fractional radius $r_i= r/R_\odot$ (0.1, 0.15, and 0.2) which is equal to $\Omega_\mathrm{rad}$=433 nHz. From each of this internal fractional radius ($r_i$) down to the center we have a step-like profile having a rate of three, five, and ten times larger than the rest of the radiative zone. Finally, we have also computed a rigid rotation rate down to the center of the Sun which is considered as our reference rotation profile.

\section{Methodology}

To compare the signature of the $g$-mode asymptotic properties between real and modeled spectra or even between two different synthetic spectra, we have used the method developed in \inlinecite{2007Sci...316.1591G}.  Indeed, a numerical simulation of the $g$-mode power spectrum has been used (see Section S.2 of the on-line material of \inlinecite{2007Sci...316.1591G} for a full description of the simulation). Briefly, the central frequencies of the $g$ modes have been computed as described in Section 2. Their amplitudes have been considered all the same following the calculations done by \inlinecite{KumQua1996} where it has been shown that all the modes in the studied regions have roughly the same amplitudes. 
We have also simulated the $g$-mode amplitudes following different laws (\emph{e.g.} a linear variation,  a parabolic decreasing with frequency...). In all the cases, the results were qualitatively the same.
Finally one-bin line widths have been used but we have checked that the results do not significantly change when wider modes are simulated. Since we are only interested in the power spectrum, the mode-phase information is lost. It is also important to note that all the computed simulations do not contain any noise because otherwise we could not quantify between the effects coming from the models and those coming from the different noise realizations. Therefore, this study represents what we could obtain in the best conditions. The addition of noise (both instrumental and solar) would affect the data and could probably degrade the results. 

\begin{figure}[htbp*]
\begin{center}
\includegraphics[width=0.9\textwidth,clip=]{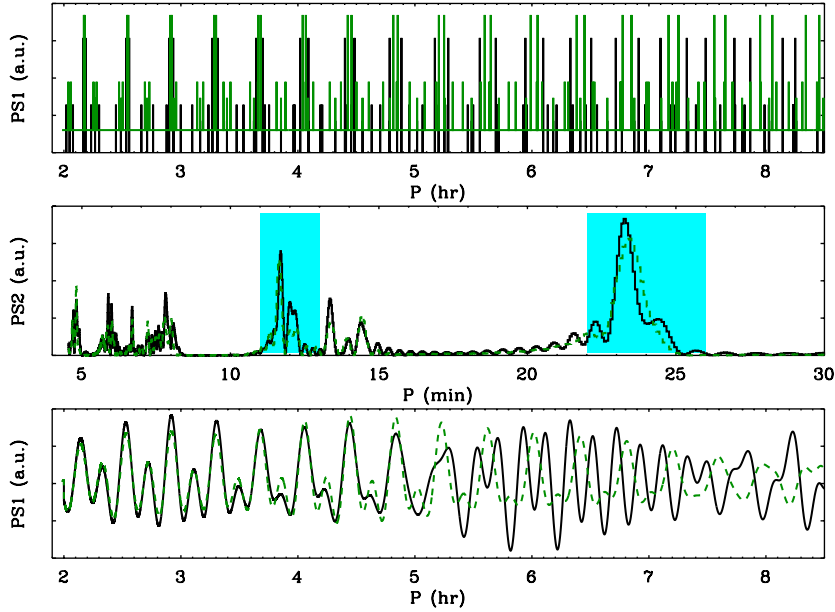}
\caption{Top panel: Power spectral density, PS1, of $\ell \,=\,1$ and 2 modes computed using the seismic model and two rotation laws in the core: a rigid core (shifted upwards in green) and a core rotating five times faster than the rest of the radiative zone at $r_i$=0.15  (continuous black lines in all the panels). Middle panel: Power spectral density, PS2, computed from PS1 between 2 and 11 hours. The shaded regions are those used to reconstruct the signal in PS1. Bottom panel: Reconstructed signals in PS1. In the two lower panels, the green dashed lines correspond to the case of a rigid core rotation law.}
\label{ps1y2}
\end{center}
\end{figure}

The starting point of this analysis is the oscillation power spectrum density. We will call it hereafter the \emph{first} spectrum (PS1). The real first spectrum is computed from GOLF time series calibrated into velocity \cite{GarSTC2005} with a classical periodogram; the synthetic ones are directly built from the $g$-mode frequencies and rotational splittings derived from the considered models (see top panel of Figure~\ref{ps1y2}).

The PS1 is expressed as a function of the period, since the $g$ modes present a comb-like structure in period and not in frequency as it is the case for $p$ modes. We have then performed a spectral analysis of the first spectrum, limited to the range 2--11 hours (\emph{i.e.} 25\,--\,140\, $\mathrm{\mu Hz}$), to get the \emph{second} spectrum (hereafter called PS2). It is important to note that the considered PS2 is a Complex spectrum, to keep information on both amplitude and phase. For practical purposes, as PS1 is not regularly sampled in period, we cannot use a classical Fast Fourier Transform (FFT), we use instead a general method based on sine-wave fittings.

The asymptotic signature of the $\ell$ = 1 $g$ modes is detected in $|\mathrm{PS2}|^2$ as a high and broad peak around 22 to 25 minutes. Depending on the physics contained on the solar model and on the rotation rate in the radiative region, the shape and the position of the $\Delta P_1$ pattern is modified. Thus in the middle panel of Figure~\ref{ps1y2} we show the resultant PS2 of the same seismic model but using two different rotation rates in the core, a rigid rotation in the whole radiative zone and a rotation rate five times faster than the rest of the radiative zone starting at $r_i\,=\,0.15$. The model of the higher rotation rate has more power at 24.5 minutes and widens the $\Delta P_1$ structure.

Once this signature (a pattern of peaks) has been detected in the PS2, we can locate in PS1 what contributes to the emergence of the structure in the PS2. To do so, we select in PS2 the vicinity around the pattern at $\Delta P_1$ and around its first harmonic at $\frac{1}{2}\Delta P_1$. We reconstruct the sum of the sine waves corresponding to the amplitudes, frequencies and phases measured in PS2 in these limited regions. In other words, we perform an inverse Fourier transform of the PS2 after having applied a narrow-band filter around what we consider to be the $\ell \,=\,1$ $g$-mode signature and its first harmonic. 

The maxima of these reconstructed waves indicate the location in PS1 corresponding to the positions where we found peaks that have a repetitive pattern. If this signature is indeed the one of the dipolar $g$ modes, the maxima of these reconstructed waves correspond to the $m$-components -- depending on their splittings -- of the $\ell$=1 $g$ modes.
The bottom panel of Figure~\ref{ps1y2} shows the reconstructed signals of the seismic models and the same two rotation rates in the core as in the previous example. At low periods, below five hours, the reconstructed waves are nearly the same in both cases, but at higher periods, the model with a higher rotation rate in the core follows the separation of the two $m$-components due to the higher splittings and the position of the maxima are different compared to the model using the rigid rotation profile.

Finally we need to quantify the consistency between our models and the GOLF observations. Comparing $\Delta P_1$ is not very informative because, as mentioned by \inlinecite{2007ApJ...668..594M} all of the theoretical computations of $\Delta P_1$ are in agreement to within a minute and, unfortunately, there is not enough resolution to reach such an accuracy in PS2. However, a direct comparison of the periods of the modes shows clear discrepancies between the different models (see bottom panel of Figure~\ref{ps1y2}). Thus, to compare the positions of the modes in PS1, we compute the Pearsons correlation coefficients (\emph{e.g.} see \opencite{1992nrfa.book.....P}) between the reconstructed waves of the models and the reference one (Section 4) or with the GOLF data (Section 5).
If the correlation coefficient is clearly positive, which means that the maxima of both reconstructed waves coincide generally, that is to say that in both considered \emph{first} spectra (real and synthetic or both synthetic), the $g$-mode components are overall at the same position.
If the modes in one spectrum are shifted compared to the modes in the other one, the correlation coefficient decreases to zero and even becomes negative. A large negative correlation indicates that each mode component in one spectrum is in between two mode components of the other spectrum.

The range used for the correlation analysis has been restricted between 2 and 8.5 hours (33 to 140 $\mu$Hz). The reason to reduce the low-frequency limit is twofold: on one hand, the dynamical effects on the periods are so important when high core rotation rates are taken into account that the $m$-components of different modes $\ell \,=\,1$ and consecutive orders ($n$) are mixed and the interpretation of the results is more complicated; on the other hand, the calculation of the frequencies at such rapid periods (above nine hours) could be less reliable and numerical effects can play an important role as it has been said in the previous section.

\section{Comparison with the Seismic Model}

\begin{figure}[htbp]
\centerline{\includegraphics[width=0.9\textwidth,clip=]{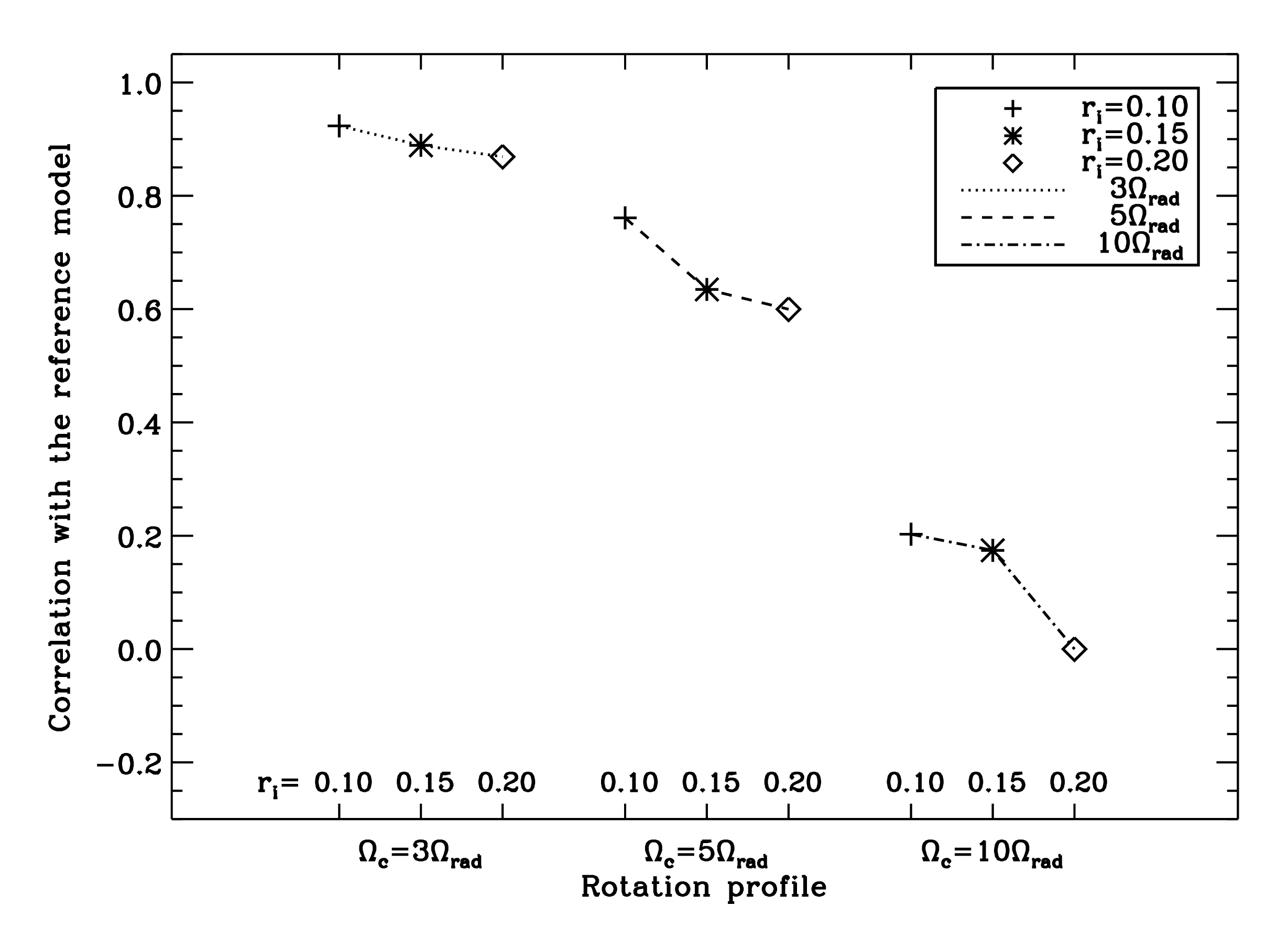}}
\caption{Correlation coefficients between the reconstructed waves for the seismic model with a rigid rotation profile and the ones for the seismic model with other rotation profiles. All of the rotation profiles have a differential rotation in the convective zone and a rigid rotation from 0.7 down to $r_i$ (= $r/R_\odot$): 0.1 (dotted line), 0.15 (dashed line), and 0.2 (dotted-dashed line). The rotation rate below $r_i$ is constant: $3\, \Omega_\mathrm{rad}$ (crosses), $5\, \Omega_\mathrm{rad}$ (asterisks), and $10\, \Omega_\mathrm{rad}$ (diamonds), where $\Omega_\mathrm{rad}$ is the rotation rate in the rest of the radiative zone (433 nHz).}
\label{correl_seis_seis}
\end{figure}

We have seen in the previous sections that the period of the $g$ modes in the low-frequency region (below 150 $\mu$Hz) is dominated by the physics and the dynamics of the radiative region, and mainly inside the solar core. We start studying the influence of the dynamics on the correlation coefficients. To do so, we apply the methodology described in the previous section to the seismic model with a rigid rotation profile that is taken as our reference model. Then, we compute the correlation coefficient between the waves reconstructed for this reference model and the ones for the seismic model with different rotation profiles. The results are shown in Figure~\ref{correl_seis_seis}. 

We can note that the higher the rotation rate in the core, the lower the value of the correlation will be (i.e. when the difference between the reference and the studied model is bigger, the correlation is smaller). For a given rotation rate in the core, if we shift the value of $r_i$ from 0.1 to 0.2 $R_\odot$, the correlation decreases a little showing that there is low sensitivity to the fractional radius $r_i$.  Then if we had only one reliable model with defined physical processes, we could infer some information on the rotation profile thanks to this correlation. We would be able to say, for instance, that a very rapid rotation in the core is very unlikely to be compatible with our reference model (seismic with rigid rotation profile).

\begin{figure}[htbp*]
\centerline{\includegraphics[width=1\textwidth,clip=]{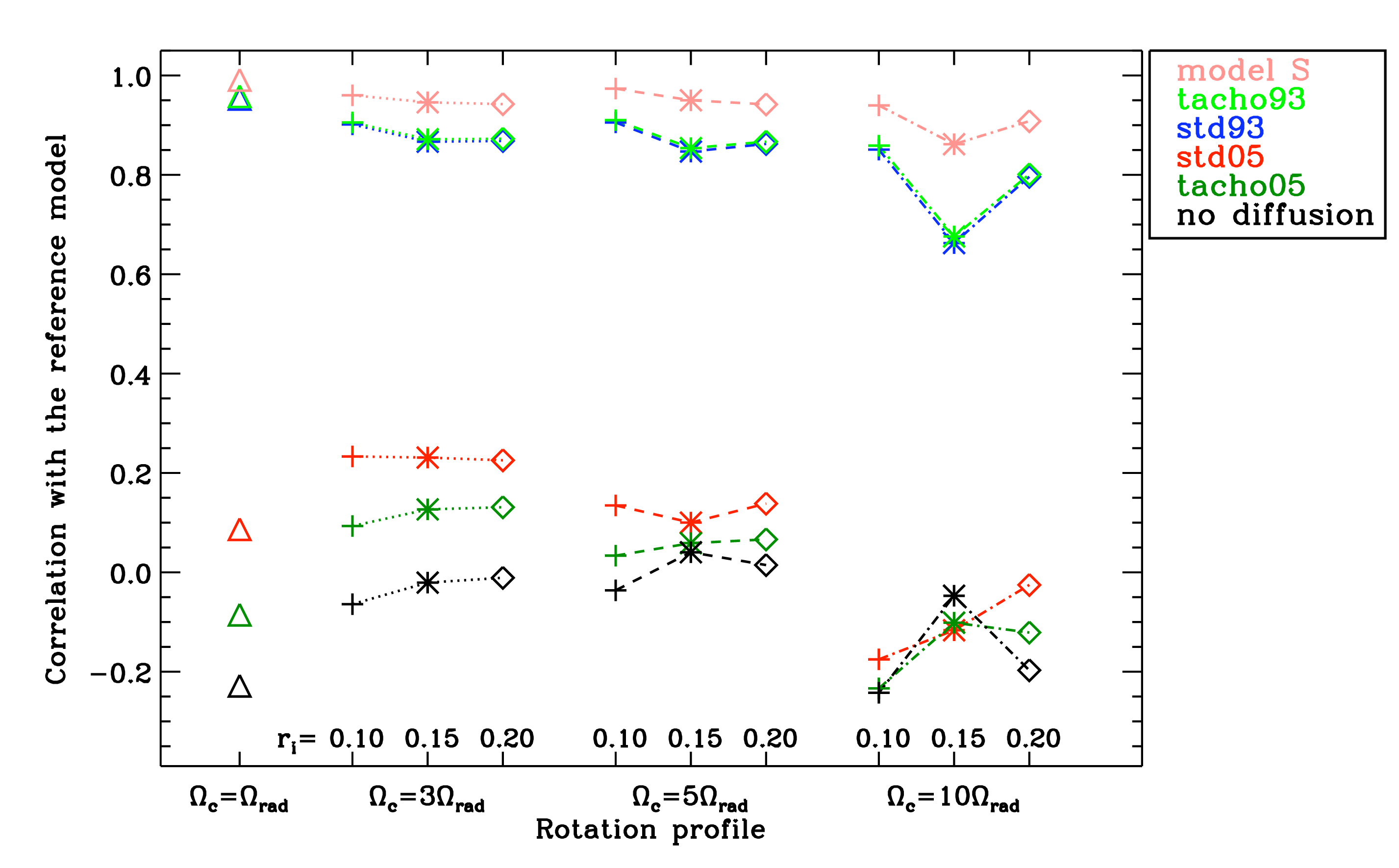}}
\caption{Correlation coefficients between the reference model (seismic) with a given rotation profile depending on the rate in the region below $r_i$ as described in Figure \ref{correl_seis_seis} (plus the rigid rotation profile (triangles)), and another model with the same rotation profile. These models (described in Section 2) are: {\it std93} (blue), {\it std05} (red), {\it tacho93} (light green), {\it tacho05} (dark green), {\it no diffusion} (black), and {\it model S} (pink).}
\label{correl_seis_mod}
\end{figure}

However, another parameter has to be taken into account: the solar structure. In Figure \ref{correl_seis_mod}, we have plotted the correlation coefficients between the reconstructed waves for the seismic model with a given rotation profile and another model with the same rotation profile. As each point on the plot is calculated with two different models but with the same rotation profile, it indicates the sensitivity of the correlation to the different physical parameters included in both compared models. 

We can see in Figure \ref{correl_seis_mod} that there are two groups of correlations clearly separated. The first one contains the models with the old abundances of \emph{GN93} that give correlations above 70\% whatever rotation profile to which they are attributed. Whereas the second group of models containing the model without diffusion and the models that include the new abundances of \emph{As05}, gives correlations below 30\% and even negative. Thus the $g$-mode frequencies extracted from these two groups of modes are not compatible among themselves. While this result is already known \cite{2007A&A...469.1145Z,2007ApJ...668..594M} it proves the ability of this methodology to separate both types of models. Obviously, there are tiny differences between other models such as the {\it std93} and the {\it tacho93} models because the frequency differences between the $g$ modes computed from them are very small. 

Moreover, we can see another systematics in Figure~\ref{correl_seis_mod} as the correlation coefficient decreases with higher rotation rates in the core. When higher rotation rates are considered, the reconstructed waves have to match both $m \pm \,1$ components of the $\ell \,=\,1$ modes. Therefore, the structure of the waves is more complicated allowing a higher precision in the calculation of the correlation and more sensitivity to the positions of the modes (the structure parameters of the model).

\section{Comparison with the Real Data}

\begin{figure}[htbp*]
\centerline{\includegraphics[width=1\textwidth,clip=]{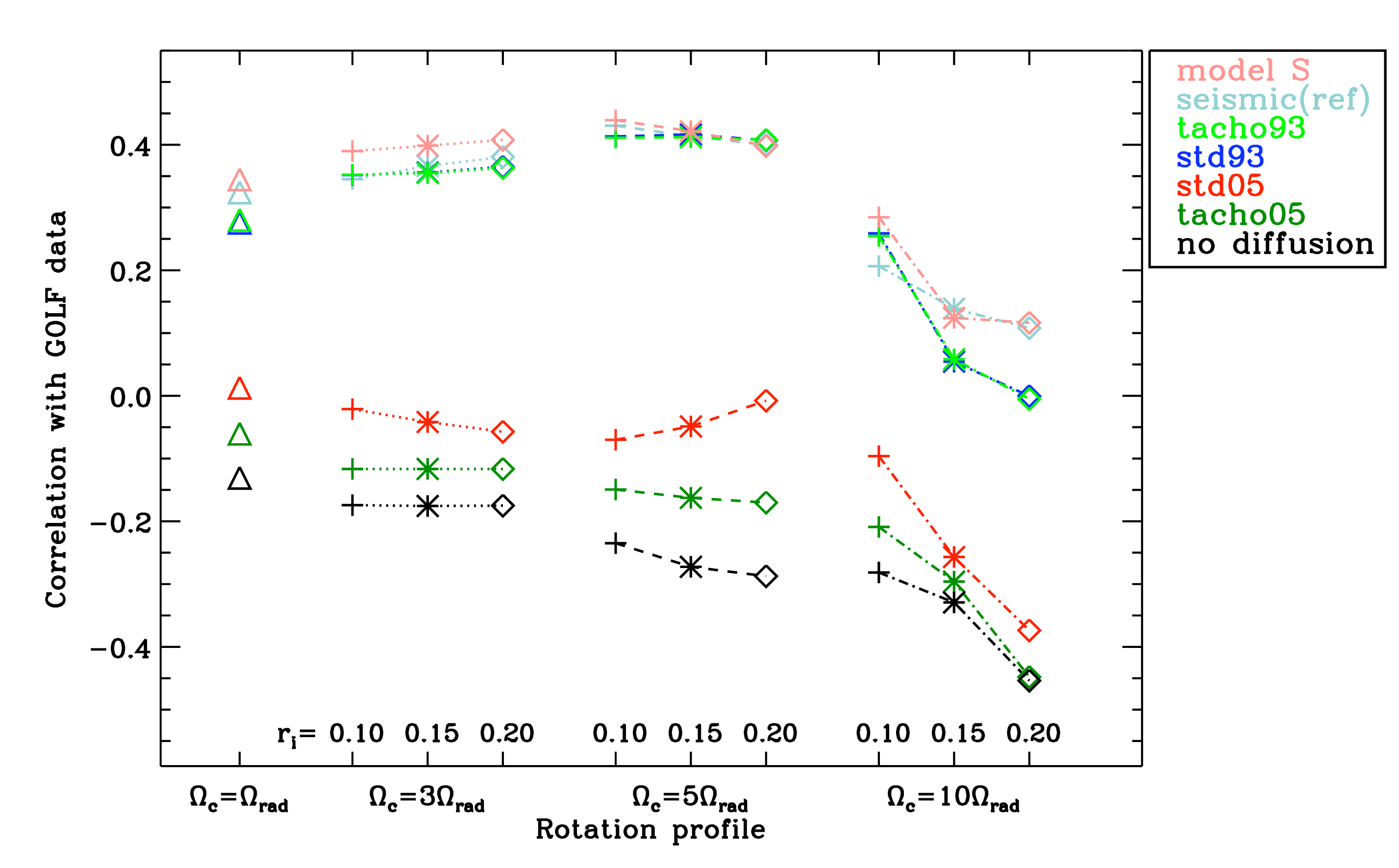}}
\caption{Correlation coefficients between the reconstructed waves from GOLF data and the ones from all of the models with the various rotation profiles. These models (described in Section 2) are: {\it std93} (blue), {\it std05} (red), {\it tacho93} (light green), {\it tacho05} (dark green), {\it no diffusion} (black), {\it model S} (pink), and {\it seismic} (light blue).}
\label{correl_golf_mod}
\end{figure}

To compare our models with the real data, we use 3481 day-long GOLF time series that were also used in \inlinecite{2007Sci...316.1591G} (see that paper for more details on the data analysis process). Figure~\ref{correl_golf_mod} shows the results of the correlation between the reconstructed waves obtained from the GOLF data and those from the models. In this figure, instead of having one free parameter, (physical processes or rotation), the results depend on both structure and dynamics. However, by mixing them, it is possible to extract some information on the data.

As in the case of the correlation with the reference model, we see a clear separation between the same two groups of models: on one side we have all the models using \emph{GN93} opacities and, on the other, the models using \emph{As05} opacities and the model without diffusion. This was an expected result. Indeed, we knew that GOLF is well correlated with the seismic model for rotation rates in the core not faster than five times -- on average -- the rotation rate in the rest of the radiative region whatever the considered fractional radius ($r_i$). As we have seen in the previous section, the models containing the new abundances and the one without the microscopic diffusion are not compatible with the models including the old chemical abundances. Therefore, they could not be compatible with the GOLF data. Actually, models that are close to the seismic model (that is to say, with the \emph{GN93} composition) associated with a rotation rate in the core $\Omega_c \le 5\, \Omega_\mathrm{rad}$ present the best compatibility with the data. A rotation profile having a rate five times the rotation rate in the rest of the radiative zone, gives the highest correlation coefficients.

It is also important to note that the models using \emph{As05} and the one without the treatment of the microscopic diffusion have, in most cases, negative correlations. The $g$-mode frequencies of these models are shifted as compared to GOLF and the maxima of the reconstructed waves are placed in between those of the GOLF data. 

We know that the largest difference of $g$-mode frequencies can reach up to 5 $\mu$Hz between the models with the \emph{GN93} abundances and the other group of models. In contrast to that, among the group of models presenting the highest correlations in Figure \ref{correl_golf_mod}, the frequencies are shifted by less than 1 $\mu$Hz (see \opencite{2007ApJ...668..594M}). Therefore, we can say that using this technique on GOLF data, we are able to distinguish between structural parameters of the solar models (such as the abundances) when the frequency differences induced by these modifications are of the order of 1 $\mu$Hz or even larger. If solar models that differ by various physical processes, produce $g$-mode frequencies with differences lower than a $\mu$Hz, the present methodology would not be able to distinguish between them. To improve our detection capability we should enlarge the analyzed region toward lower frequencies where we have seen that the $g$-mode periods are sensitive to the models.

\section{Conclusion}

In the present paper, we have studied the changes in the signal reconstructed from the period separation $\Delta P_1$, which vary when the physics of the solar models are modified, and we have verified whether it was possible to detect these changes in the $g$-mode asymptotic signature with the present uncertainties. 

By comparing our different models with the reference one, we have learned that the methodology presented here offers enough sensitivity to distinguish between two sets of models: first, diffusive models using the old \emph{GN93} abundances; second, models using the new \emph{As05} abundances and the one not including microscopic diffusion.

This work shows that the $g$-mode signature detected in GOLF data clearly
favours the models including the old abundances from \inlinecite{1993oee..conf...14G}. In this sense, the analysis of the $g$-mode signature recently detected is in agreement with previous results coming from the $p$-mode analysis showing a larger discrepancy in the sound-speed profile when we use the new abundances from  \inlinecite{2005ASPC..336...25A}.

However, we have to keep in mind that all of these results do not mean that the new abundances are incorrect. They only tell us that the classic models including them are not compatible with the observations. But it could be possible to obtain a better agreement by changing other physical quantities of the models such as the opacities or the reaction rates or by including new processes to compensate for the effects of the reduction in the metallicity.

\begin{acks}
The authors want to thank J. Christensen-Dalsgaard who provided us
the model S and the adipack code, J. Provost for having given us her frequencies that we could compare with the ones of our standard model and for having produced the seismic-model frequencies with her own oscillation code, and A. Eff-Darwich and S. Turck-Chi\`eze for useful discussions and comments. The GOLF experiment is based upon a consortium of institutes (IAS, CEA/Saclay, Nice, and Bordeaux Observatories from France, and IAC from Spain) involving a large number of scientists and engineers, as enumerated in \inlinecite{GabGre1995}. SOHO is a mission of international cooperation between ESA and NASA. This work has been partially funded by the grant AYA2004-04462 of the Spanish Ministry of Education and Culture and by the European Helio- and Asteroseismology Network (HELAS: {\url{http://www.helas-eu.org/}}), a major international collaboration funded by the European Commission's Sixth Framework Programme.
\end{acks}

   
\bibliographystyle{spr-mp-sola}

\bibliography{/Users/rgarcia/Desktop/BIBLIO}  

\IfFileExists{\jobname.bbl}{} {\typeout{}
\typeout{****************************************************}
\typeout{****************************************************}
\typeout{** Please run "bibtex \jobname" to obtain} \typeout{**
the bibliography and then re-run LaTeX} \typeout{** twice to fix
the references !}
\typeout{****************************************************}
\typeout{****************************************************}
\typeout{}}

\end{article} 
\end{document}